\documentclass[aps,nofootinbib,superscriptaddress,tightenlines,twocolumn,pra,longbibliography]{revtex4-1}
\usepackage{amsmath,amssymb,amstext,amsthm}
\usepackage[pdftex]{graphicx}
\usepackage{braket}
\usepackage{bbm}
\usepackage{xcolor}
\definecolor{dred}{rgb}{.8,0.2,.2}
\definecolor{ddred}{rgb}{.8,0.5,.5}
\definecolor{dblue}{rgb}{.2,0.2,.8}
\usepackage[pdftex,letterpaper=true,pagebackref=false]{hyperref}
\hypersetup{
pdftitle={Noise Refocusing in a Double Loop Mach-Zehnder Neutron Interferometer},
pdfauthor={J. Nsofini},
pdfsubject={Quantum information, quantum physics, neutron interferometry, quantum coherence}
	pdfnewwindow=true, 
	colorlinks=true, 
	linkcolor=blue, 
	citecolor=blue, 
	filecolor=blue, 
	urlcolor=blue
}

\def\beq{\begin{equation}}
\def\eeq{\end{equation}}
\def\bsp{\begin{split}}
\def\esp{\end{split}}
\def\bea{\begin{eqnarray}}
\def\eea{\end{eqnarray}}
\def\ba{\begin{array}}
\def\ea{\end{array}}

\graphicspath{{Figures/}}
\begin{document}

\title{ Coherence Optimization in Neutron Interferometry through Defocussing}
\author{J. Nsofini} 
\email{jnsofini@uwaterloo.ca}
\affiliation{Department of Physics, University of Waterloo, Waterloo, ON, Canada, N2L3G1}
\affiliation{Institute for Quantum Computing, University of Waterloo,  Waterloo, ON, Canada, N2L3G1}
\author{D. Sarenac}
\affiliation{Department of Physics, University of Waterloo, Waterloo, ON, Canada, N2L3G1}
\affiliation{Institute for Quantum Computing, University of Waterloo,  Waterloo, ON, Canada, N2L3G1} 
\author{D. G. Cory}
\affiliation{Institute for Quantum Computing, University of Waterloo,  Waterloo, ON, Canada, N2L3G1} 
\affiliation{Department of Chemistry, University of Waterloo, Waterloo, ON, Canada, N2L3G1}
\affiliation{Perimeter Institute for Theoretical Physics, Waterloo, ON, Canada, N2L2Y5}
\affiliation{Canadian Institute for Advanced Research, Toronto, Ontario, Canada, M5G 1Z8}
\author{D. A. Pushin}
\affiliation{Department of Physics, University of Waterloo, Waterloo, ON, Canada, N2L3G1}
\affiliation{Institute for Quantum Computing, University of Waterloo,  Waterloo, ON, Canada, N2L3G1}

\begin{abstract}

A zero-area four-blade perfect crystal neutron interferometer (NI) possess a decoherence-free subspace (DFS) for low-frequency mechanical vibrations and thus is easier to site. 
However, unlike the standard three-blade Mach-Zehnder NI the ideal contrast of this four-blade NI geometry is less than one. By applying a recently introduced quantum information model for dynamical diffraction we show that the contrast for the four-blade DFS NI can be increased by offsetting the focusing condition. The contrast optimization leads to an NI geometry where the distances between the centers of the blades are equidistant. An experiment is proposed to verify the increase in contrast.


\end{abstract}

\maketitle


\section{Introduction}
\label{sec:intro}
Perfect crystal neutron interferometry is a powerful tool 
for characterization of materials and the precise measurements of fundamental constants from condensed matter physics and Standard Model of particle physics \cite{Rauch_1975_PhysLetta,Colella_1975_PRL,Schoen_2003_PhysRev,Green_2007_PRC,Ioffe_2002_ApplPhysA,Rauch_2015_Book,oam,chameleon,holography,lemmel2015neutron}. 
However, the narrow momentum acceptance and the required stringent forms of environmental vibration isolation limit the availability of this technique \cite{Shahi_2016_NIM,Pushin_2015_AHEP,saggu2016decoupling}. 

In 2011, a unique design of a zero-area perfect crystal four-blade NI, commonly referred to as the DFS NI, was experimentally demonstrated to have a subspace which protects information from low-frequency mechanical vibrational noise~\cite{Pushin_2011_PRL}. 
Its design employs two symmetry requirements. The first stipulates that the two paths inside the NI should be refocused on the last blade
~\cite{Rauch_2015_Book}, as shown in Fig.~\ref{Fig:Laue}a.  
The second symmetry requirement is that there are two loops of equal area inside the NI so that the noise induced in the first loop is removed in the second loop~\cite{Pushin_2009_PRA}.
Its unique design overcame a significant obstacle which limits the application of perfect crystal neutron interferometry. 
In spite of this, the observed contrast of 25\% in \cite{Pushin_2011_PRL} was less than half of the expected contrast. A recent study suggests that the maximum achievable contrast for the DFS NI is unavoidably less than one~\cite{Nsofini_2017_JAP}.

In this work we show that by relaxing the first symmetry condition that demands the two paths inside the NI to refocused on the last blade, we show that the contrast increases to a maximum at a specific defocusing condition. The optimal contrast occurs when the distances between the centers of the blades are equal, as shown in Fig.~\ref{Fig:Laue}b.
Furthermore, we propose an experiment to verify the increase in contrast with the original DFS NI geometry. Our simulations rely on the recently developed quantum information (QI) formalism of dynamical diffraction (DD) which is based on the repeated application of a unitary at coarse-grained site \cite{Nsofini_2016_PhysRevA}.



 
 
This article is organized as follows: In Sec.~\ref{Sec:IntroDD} we introduce the QI model including its application to the DFS NI. In Sec.~\ref{Sec:EffectsofDef} we  compare the effects of defocussing on the DFS NI and the three-blade NI and then assess the extend to which the  noise refocusing ability is affected. In Sec.~\ref{Sec:PropoExp} we propose an experiment to verify the concepts, and in Sec.~\ref{Sec:Conclusion} are the concluding remarks.

\section{Modeling DD from a four-blade NI} \label{Sec:IntroDD}

The theory of DD takes into account  multiple coherent scattering to describes the propagation of waves through perfect periodic lattices including crystals and self-assembly systems \cite{Rauch_2015_Book,Authier_2006_Book,Utsuro_2012_Book}. 
There are excellent reviews of DD available for both x-rays and neutrons with a focus on the two-wave approximation~\cite{Zachariasen_1945_Book,Batterman_1964_RevModPhys,Authier_2006_Book,Sears_1989_Book,Utsuro_2012_Book}. For a brief overview of the standard theory of DD and its application to the DFS NI, see Appendix~\ref{AppendixDD}.
In this article, we apply the QI model of DD \cite{Nsofini_2016_PhysRevA} to analyze coherence effects in the four blade NI geometry. The advantage of this model is that it can easily and quickly simulate a wide range of geometries. We start by reviewing the application of the model to a single NI blade.

\subsection{QI model of a single blade} \label{Sec:IntroDD_oneBlade}
The QI formalism of DD is based on the discretization of the scattering media into coarse-grained units which acts as unitary operators \cite{Nsofini_2016_PhysRevA}. The action of an NI blade is also a unitary operator expressed as a combination of multiple $2\times 2$ unitary operators at coarse-grained sites (nodes). The two-level logical subsystem of a transversing wave is labeled as a state vector $|{a_j}\rangle$ or $|{b_j}\rangle$ where $a$ $(b)$  refers to the rays moving upward $k_y>0$ (downward $k_y<0$), and $j$ indexes the node. 
Because the trajectories are distinct and coherent,  the ray tracing approach is analogous to a path integral, so the QI model leads to a blade unitary operator, ${\cal U}(N)$, expressed as
\bea
{\cal U}(N)=\prod_{m=1}^N{\cal U}_m=\prod_{m=1}^N\sum_{j=i}^mU_j(\xi,\theta,\zeta)\label{Eq:BladeUnitary},
\eea
where $N$ is the number of layers of $m$-nodes each with the following unitary operator
\begin{align}
U_{j}(\xi,\theta,\zeta)
	&=  \ket{a_{j+1}}\left(e^{i \xi}\cos\theta\bra{a_j}+  e^{i \zeta}\sin\theta \bra{b_j}\right)\nonumber\\
	&-\ket{b_{j-1}}\left(e^{-i \zeta}\sin\theta\bra{a_j}-  e^{-i \xi}\cos\theta\bra{b_j}\right), \label{Eq:coinOp}
\end{align}
\noindent where $\xi$ $(\zeta)$ is the phase of the transmitted (reflected) beam of a single node, and $\theta$ controls the relative probability amplitude of the reflected and transmitted beams of a single node. 

In order to extend the QI model to an NI geometry, we consider a single path entering a single crystal blade:
\begin{align}
\ket{\Psi^\text{in}}=\ket{a_0},
\end{align}
on the node $j=0$. Upon interaction with a single crystal with operator ${\cal U}(N)$, the state at the exit is:
\begin{align}
\ket{\Psi^\text{out}}=\sum_j\Big(\alpha_j\ket{a_j}+\beta_j\ket{b_j}\Big),
\end{align}
where,  $\alpha_j=\bra{a_j}{\cal U}(N)\ket{\Psi^\text{in}}$ and $\beta_j=\bra{b_j}{\cal U}(N)\ket{\Psi^\text{in}}$ are the probability amplitude coefficients, similar to Eqs.~(\ref{Eq:DDOneStandard} \& \ref{ddeq2}) derived in the appendix  using the standard theory of DD.

\begin{figure}[!t]
\center
\includegraphics[width=0.99\columnwidth]{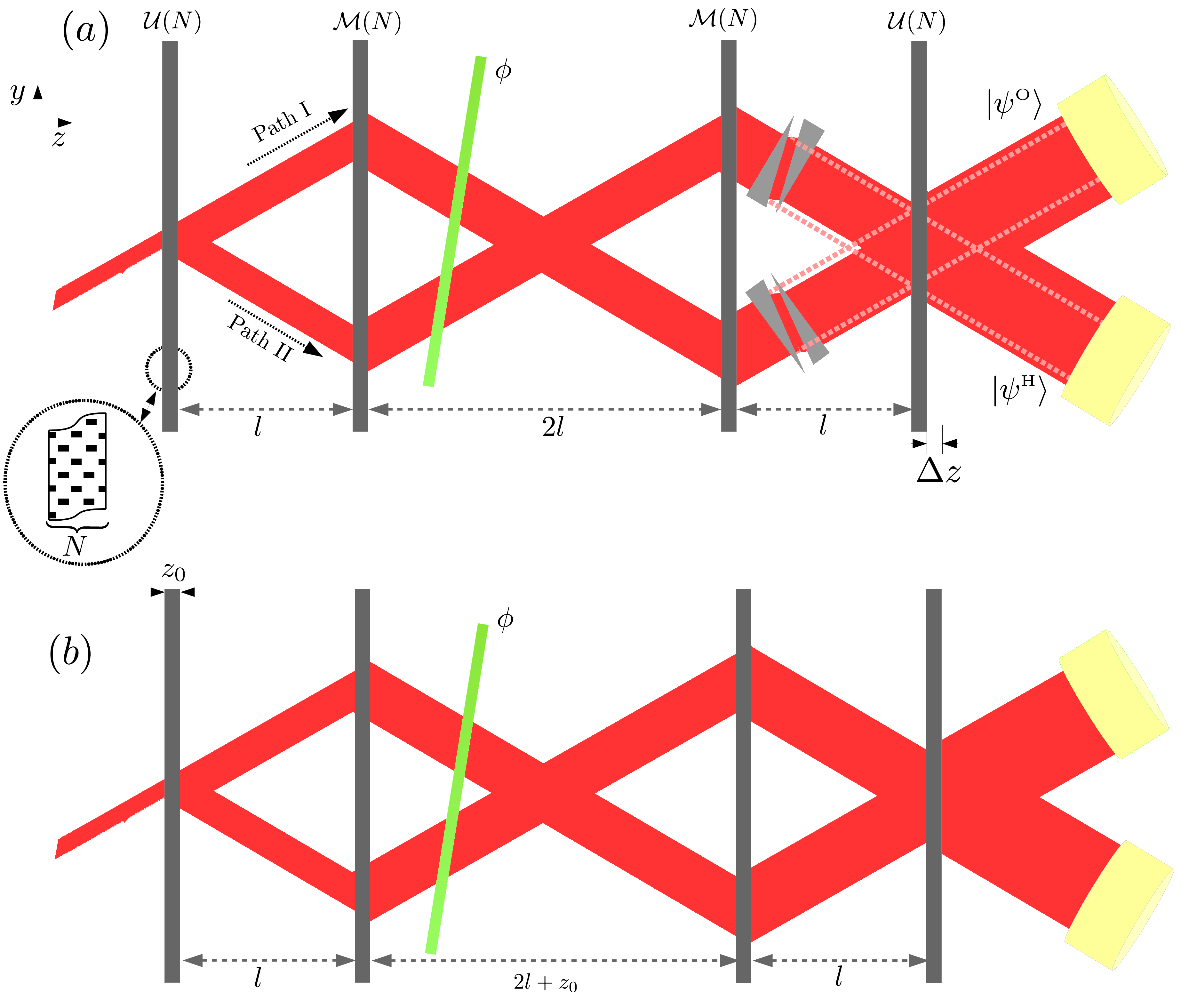}
\caption{
(a) A sketch of a four-blade DFS NI geometry where on the last blade the two coherent beams are defocussed using pairs of prisms. When the prisms are taken out, the beams overlap on the third blade such that the areas of the two interferometer loops are the same. b) In this geometry the distances between the centers of the blades are equal, and the area enclosed in the first loop is slightly different from that of the second loop. This geometry is found to be optimal for contrast.
} \label{Fig:Laue}
\end{figure}

\subsection{Modeling the Four-blade DFS NI}
\label{sec:FOurbladeNI}

Here we describe how to model the four-blade DFS NI with the QI model of DD. A detailed description of the four-blade NI geometry can be found in ref.~\cite{Pushin_2009_PRA}. It consists of four identical crystal blades sitting on a common base, see Fig.~\ref{Fig:Laue}a. 
When a beam is incident on the first blade, it splits into a coherent superposition of two components that are each redirected by the second and third blades onto the fourth blade where they recombine and interfere. 
Although, each of the NI blades act as a beam splitter, a post selection on only those neutrons that reach the detectors ensure that the second and third blade act as mirrors~\cite{pushin2006coherent}. 


The four-blade DFS NI shares features with the three-blade Mach-Zehnder (MZ) interferometer, which is the commonly studied geometry of neutron interferometry.  
On a general note the major difference is that the two paths are redirected twice before reaching the last blade.
This distinctive feature enables the four-blade NI to posses a DFS for low-frequency mechanical vibrational noise \cite{Pushin_2011_PRL}. 

In the four-blade DFS NI shown in Fig.~\ref{Fig:Laue}a, the  QI formalism of DD stipulates a unitary operator for each of the first and last blades, here denoted by  ${\cal U}(N)$. The operator of the middle two blades is derived from ${\cal U}(N)$ using appropriate projection operators.
The projection operators are defined on the upward propagating beam (O-beam) and the downward propagating beam (H-beam) as
\begin{align}
P^\text{\tiny O}=\sum_j\ket{a_j}\bra{a_j},\quad P^\text{\tiny H}=\sum_j\ket{b_j}\bra{b_j}.
\end{align}
We can then express the nonunitary operator of the middle blades which act as mirrors as follows
\begin{align}
{\cal M}(N)= P^\text{\tiny H}{\cal U}(N)P^\text{\tiny O} +P^\text{\tiny O}{\cal U}(N)P^\text{\tiny H}.
\end{align}
%

If we denote by $\Phi(\phi)$ the operator  which induces a phase shift of $\phi$ between the two paths, then the state at the output of the interferometer can be expressed as
\begin{align}
\ket{\Psi^\text{\tiny NI}} 
=U^\text{\tiny NI}(N,\phi)\ket{\Psi^\text{in}}
=\sum_j\Big(\Psi^\text{\tiny O}_j\ket{a_j}+\Psi^\text{\tiny H}_j\ket{b_j}\!\Big),
\end{align}
with $U^\text{\tiny NI}(N,\phi)\!=\!{\cal U}(N){\cal M}(N){\cal M}(N)\Phi(\phi) {\cal U}(N)$, and the O-beam component labeled $\Psi^\text{\tiny O}_j\!=\!\bra{a_j}U^\text{\tiny NI}(N,\phi)\ket{\Psi^\text{in}}$  and  the H-beam component labeled $\Psi^\text{\tiny H}_j\!=\!\bra{b_j}U^\text{\tiny NI}(N,\phi)\ket{\Psi^\text{in}}$ are functions of $\phi$. 
%
The intensities at the O-beam and H-beam are sinusoidal
\begin{align}
I^\text{\tiny H}&= \mathcal{A}^\text{\tiny H}+\mathcal{B}^\text{\tiny H}\cos\left(\phi-\varphi\right),\label{Eq:OHQIModel1}\\
I^\text{\tiny O}&=  \mathcal{A}^\text{\tiny O}+\mathcal{B}^\text{\tiny O}\cos\left(\phi-\varphi\right), \label{Eq:OHQIModel}
\end{align}
 where the coefficients, ${\cal{A}}^\text{\tiny O}$ and ${\cal{A}}^\text{\tiny H}$ are the constant part of the intensity functions, and $\mathcal{B}^\text{\tiny O}$ and ${\cal{B}}^\text{\tiny H}$ are the amplitudes of the oscillating part of the respective intensities. From conservation of neutron counts reaching the detectors, $I_H+I_O=constant$, it follows that $\mathcal{B}^\text{\tiny O}=-{\cal{B}}^\text{\tiny H}$. The phase accumulated from the operators is $ \varphi$ and will be described later.
A similar intensity behavior is expected when the standard theory of DD is used, see  Eq.~(\ref{Eq:FourBlade-DD}) in Appendix A for details.

The measure of the quality of the interference pattern at the the exit of the interferometer is called the contrast and it is expressed by
\begin{align}
\mathcal{V}=\frac{I_\text{max}-I_\text{min}}{I_\text{max}+I_\text{min}},
\label{Eq:contrast}
\end{align}
where $I_\text{max}$ and $I_\text{min}$ are the maximum and minimum intensities at the particular detector. 
Considering the H-beam, the contrast $\mathcal{V}=|\mathcal{B}^\text{\tiny H}/\mathcal{A}^\text{\tiny H}|$, is then a function of the parameters $\xi,\theta,\zeta$ and $N$. Therefore, a choice of the parameters must be made for any NI geometry. An efficient way to do this is by performing an optimization  of the contrast over these parameters. In our case, after setting $\xi=0$ and $\zeta=0$, we obtained an optimal contrast of 0.68 at ${N} = 44$ and $\theta=\pi/64$ for a DFS NI.

\section{Effects of Defocussing}\label{Sec:EffectsofDef}

In the four-blade DFS NI geometry the separation between the first and second blades is equal to the distance between the third and fourth blades, while the distance between the second and third blade is twice as long.  The fact that the separation between the two middle blades is twice as long as the other blade separations leads to a breaking of translational invariance of the blades. That is, the blades are not positioned at equal distances in this geometry. 
The cost of breaking this symmetry  has been shown recently to cause a decrease in the contrast of the four-blade NI \cite{Nsofini_2017_JAP}.
Stated in another way, the phases occurring due to diffraction inside the blade are not equal for Path I and Path II in this NI geometry. 
The relative phases in the two paths are spatially offset as follows:

\begin{align}
\phi_\text{II}=\phi_\text{I}+\Delta\phi_m
\end{align}

In this work we observe that displacing the last blade by an amount equal to the thickness of the blade improves the contrast. 
By moving the last blade on which the neutron wavefunction recombines, an extra phase is induced which cancels out (refocuses) the phase difference between the two paths.

\begin{figure}[!t]
\center
\includegraphics[width=.95\columnwidth]{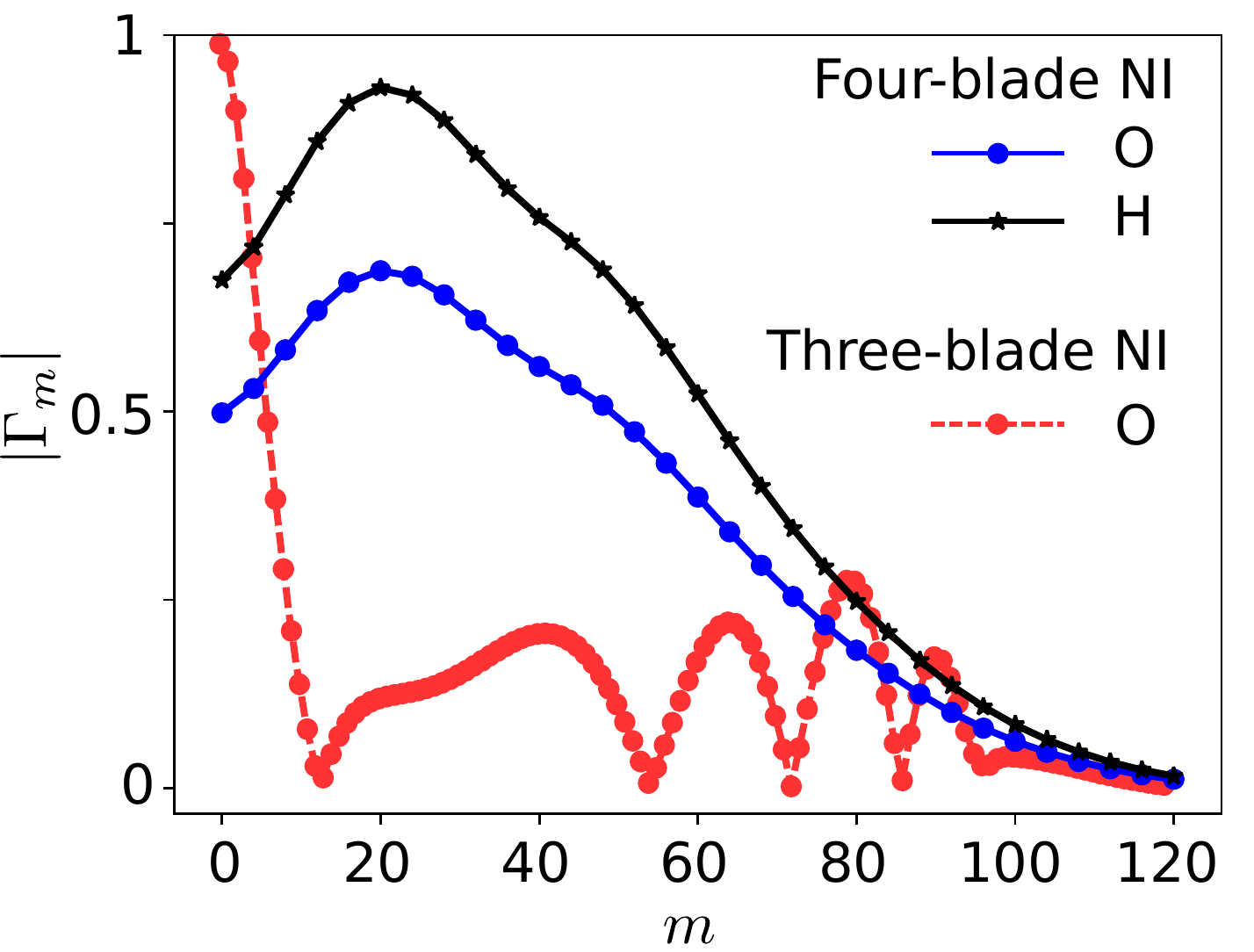}
\caption{Simulations of the contrast obtained using the QI model as a function of the defocussing parameter $m$. 
The contrast of the four-blade DFS NI increases to a maximum at $m=N/2$, for ${N} = 44$ and $\theta=\pi/64$, and then decreases. It can also be noted that the contrast of the H-beam is higher than that of the O-beam as expected~\cite{Pushin_2009_PRA}.
(red curve) The contrast of the three-blade NI (with ${N} = 100$ and $\theta=\pi/32$) starts at a maximum value and then decrease to zero with a good agreement to that shown in Ref.~\cite{PETRASCHECK_1988_PhysicaB}.
} \label{Fig:Coherence4BNI-QIDD}
\end{figure}

One way of modeling defocussing (translating the last blade by $\Delta z$ along the z direction in Fig.~\ref{Fig:Laue}) using the QI model is by translating the neutron wavefunction of one path by $m$ nodes in the transverse direction (y direction in Fig.~\ref{Fig:Laue}). A beam mismatched by $m$ is equivalent to that defocused by $\Delta z$. We can then introduce a coherence function $\Gamma_m$ quantifying the effect of defocussing:

\bea
\Gamma_m=\sum_j\braket{\psi_j^\text{II}|\psi_{j+m}^{\text{I}}}\label{Eq:CoherentFQI},
\eea
\noindent where $\ket{\psi_{j+m}^{\text{I}}}$ and $\ket{\psi_j^\text{II}}$ are the contribution to the wavefunction of path I and path II at node $j$, and where we increment the wavefunction of path I by m nodes. In connection to Eqs.(\ref{Eq:OHQIModel1},\ref{Eq:OHQIModel}), the quantities $|{\cal{B}}_\text{\tiny H}|$ and $\varphi_\text{\tiny H}$ are the absolute value and the argument of the coherence function. Note that $\Gamma_m$ is analogous to the coherence function defined in the standard theory of DD, see Eq.~(\ref{Eq:CohFunction-SDD}) in Appendix.

Fig.~\ref{Fig:Coherence4BNI-QIDD} represents a plot of the absolute value of the coherence function (which is also equal to the contrast) as a function of the  defocussing parameter $m$. 
The plot shows that by displacing the last blade, or equivalently increasing $m$, the contrast increases to a maximum and then starts to decrease again. 
The maximum contrast occurs at a blade displacement $\Delta z$ which equals the crystal thickness $z_0$ or equivalently $N$. This displacement corresponds to $m=N/2$  for the optimal values ${N} = 44$ and $\theta=\pi/64$.
Similar results can be reproduced from the standard theory of DD, which is covered in the appendix. 
Both models confirm that by introducing a defocussing, a four-blade NI can be engineered to have a higher contrast in the expense of an asymmetry in the two loops. 

Introducing a defocusing in a standard three-blade NI has been  extensively studied, especially in the determination of the longitudinal and transverse coherent lengths of a neutron wavepacket \cite{Rauch_2015_Book,Petrascheck_1976_PhysStatSol,Petrascheck_1984_ActaCrys,PETRASCHECK_1988_PhysicaB}. 
By displacing the beam in the longitudinal or transverse position, these experiments measured the contrast as a function of beam displacement. In a similar way we can apply the QI model to the three-blade NI and offset the beam intersecting on the surface of the third blade by $m$ nodes. The coherence function as a function of $m$, obtained for a  coarse-grained parameters $N=100$  and $\theta=\pi/32$, is shown in Fig.~\ref{Fig:Coherence4BNI-QIDD}. 
As expected we observe that it starts at a maximum (when $m=0$) and falls steadily to zero, as the NI is defocussed ($m\neq0$). The simulation shows agreement with those already presented in earlier theoretical studies of the coherence length ref.~\cite{PETRASCHECK_1988_PhysicaB} and  observed experimental measurement of the transverse and longitudinal coherent lengths~\cite{Rauch_2015_Book,Pushin_2008_PRL}.  

\begin{figure}[!t]
\center
\includegraphics[width=.99\columnwidth]{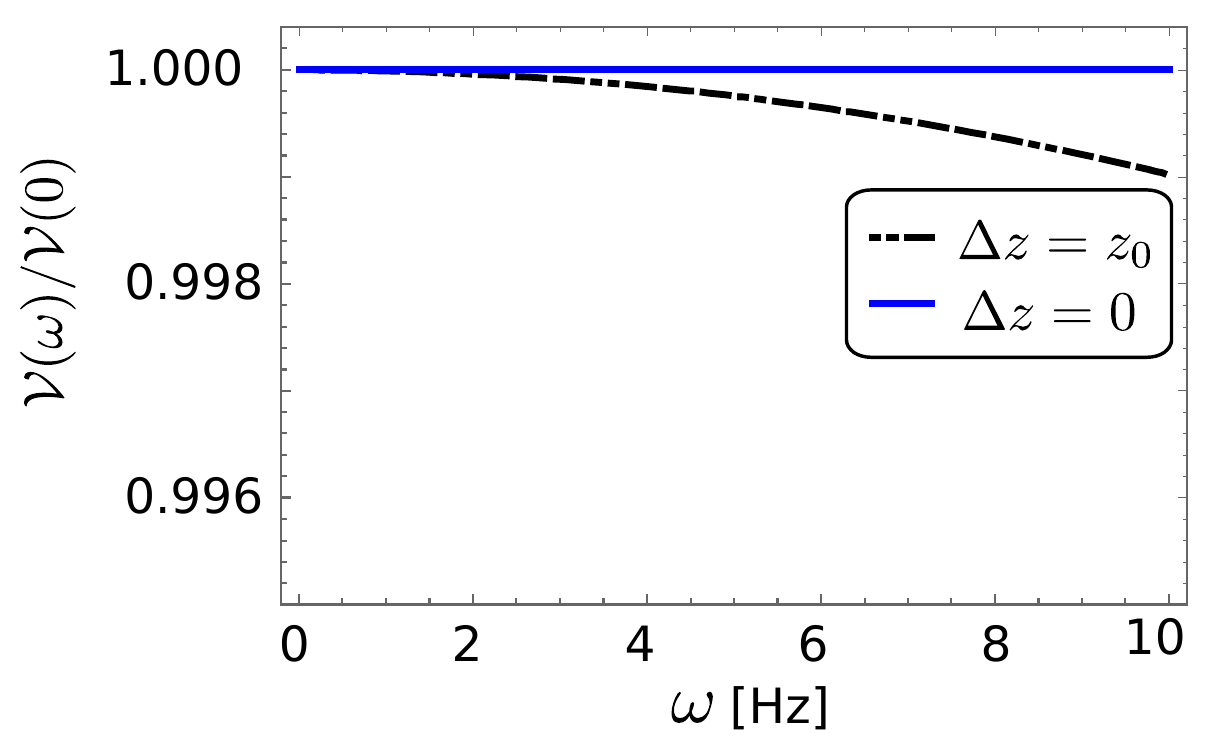}
\caption{The relative contrast in the four-blade DFS NI as a function of mechanical vibrational noise frequency. The thickness of the blades is $z_0$.  A comparison of the DFS interferometer ($\Delta z=0$) and the optimally defocussed geometry ($\Delta z=z_0$) when operating in an environment with mechanical vibrational around an axis through the center of mass and normal to the plane of Fig.~\ref{Fig:Laue}.} \label{Fig:DefocusedContrastMechVib4BNI}
\end{figure}

Although, a defocussing in the four-blade DFS NI geometry can lead to an increase in the contrast, it also breaks the symmetry of the NI. That is, the area enclosed by the first loop is different from the area enclosed by the second loop. 
This area mismatch means that a defocussed DFS interferometer might now suffer from mechanical vibrations noise. 

Here we consider only mechanical vibrational noise that originate from rotations of frequency $\omega$ around the center of mass and along the axis normal to the plane of Fig.~\ref{Fig:Laue} as they are the major contributors to contrast loss~\cite{Pushin_2009_PRA,Pushin_2011_PRL}. 
When the four-blade DFS NI  is subject to such noise, the relative contrast defined as the contrast in the presence of noise scaled by that in the absence of no noise is given by 
\begin{align}
\mathcal{V}(\omega)/\mathcal{V}(0)=|J_0(\omega^2\ {48m_nv_y v_{z}\theta_0\tau}/{\hbar})|, \label{eqn:con4}
\end{align}
where $J_0()$ is the Bessel function of the first kind, $m_n$ is the neutron mass,  $\mathbf{v}={v}_y \hat{y}+v_z \hat{z}$ is the velocity of the neutron, $\theta_0$  is the amplitude of the the noise, and $\tau$ the time a neutron of wavelength 4.4 {\AA} takes between the first and second blades. Shown in Fig.~\ref{Fig:DefocusedContrastMechVib4BNI} is the plot of the contrast eq.~(\ref{eqn:con4}) as a function of the frequency of the noise for $\Delta z=0$. 
Similar but slightly complicated expression for the contrast can be derived for the defocused four-blade NI. A simulation of this contrast is also shown in Fig.~\ref{Fig:DefocusedContrastMechVib4BNI} as a function of frequency for displacement $\Delta z=z_0$. The simulations show that the decrease in the contrast is negligible.

\section{Proposed Experiment}\label{Sec:PropoExp}

\begin{figure}[!t]
\center
\includegraphics[width=\columnwidth]{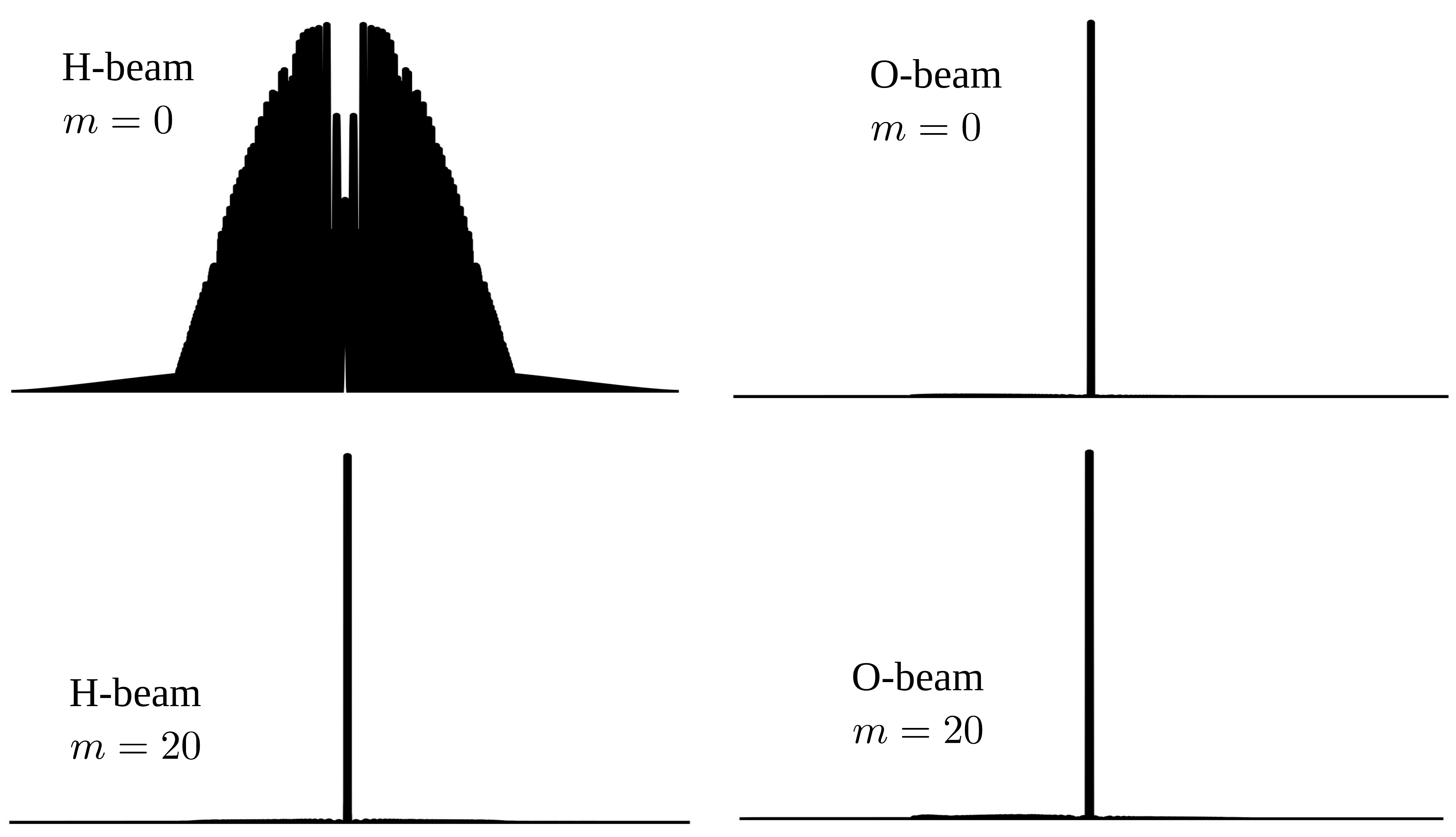}
\caption{The intensity profiles at the O-beam and H-beam with no defocussing ($m=0$) and with optical defocussing ($m=20$). The QI model parameters for this simulation were $N=20000$, and $\theta=\pi/64$. 
} \label{Fig:4BNIOHProfile}
\end{figure}

An experiment to verify that defocussing in the four-blade DFS NI leads to an increase in contrast can be achieved by detaching the fourth blade of the DFS NI and moving it by an amount $\Delta z$. However, because of the challenges that results from aligning a perfect crystal NI to an angstrom precision, it is extremely difficult to realized this. Rather we propose an alternative approach to vary defocussing in the four-blade DFS NI that will result in an increase of contrast, see Fig.~\ref{Fig:Laue}a. By displacing the beams as shown produces the same outcome as moving the last blade. 
This procedure is rather much more efficient because the experiment can be performed with the current neutron interferometry facility at the National Institute of Standards and Technology, Maryland, and using the same four-blade DFS NI that was used in prove of concept \cite{Pushin_2015_AHEP}. 
Several methods can be used to displace the beam inside the NI . For example, when a neutron beam propagates through a pair of prisms, made from a non-absorbing  uniform density  material, it is displaced by an amount determined by the refractive index of the prism material. By positioning two pairs of prism, one in each of the arms of the interferometer, the beam can be displaced by an equal amount causing the beams to be defocussed on the third blade. 
The situation in this case is exactly the same as that when the blade is moved. It should be noted that with the currently available optical elements the beam can only be displaced by a few micrometers. Therefore such a method would serve as a verification of the defocussing hypothesis and as motivation to construct defocussed four-blade DFS NI geometries. 



 In addition to measuring the contrast as a function of beam displacement, we can look at the behavior of the intensity profiles of the O-beam and H-beam. 
Shown in column 1 of Fig.~\ref{Fig:4BNIOHProfile} is the position sensitive intensity at the H-beam without a defocussing ($m=0$) and with a defocussing ($m=20$) in a four-blade NI. The coarse-grained parameters are $N=20000$ for the thickness and $\theta=\pi/64$. 
Shown in column 2 of Fig.~\ref{Fig:4BNIOHProfile} is the same plot but for the O-beam. Based on the H-beam intensities in column 1, it is clear that defocussing has a drastic effect on the intensity profile. That is, the profile with $m=20$ is centralized around the origin while that with $m=0$ is widely distributed. Note that the heights of the figures are not to scale as our purpose is to illustrate the width and not the area.

\section{Conclusion} \label{Sec:Conclusion}

By applying the QI model to dynamical diffraction to formulate the coherence function we show that an optimal design for a four-blade DFS NI with contrast close to one can be achieved  by defocussing the last blade. 
The optimal contrast is found to be when the last blade is displaced by an amount equal to the thickness of the blade. By applying the same procedure to the three-blade NI, we reproduced results that have previously been observed, illustrating that the QI model is a simplistic yet efficient to simulate dynamical diffraction effects in neutron interferometry. 
We also proposed an experimental procedure to demonstrate this prediction. Although the breaking of symmetry makes the interferometer vulnerable to mechanical vibrational noise, we also showed that the effect of mechanical noise on the defocused interferometer is negligible.
Finally, we have also simulated the spatial intensity profiles for a defocussed and non-defocussed NI which show that defocussing centralizes the intensity at the output.


\section{Acknowledgements}
This work was supported by the Canada Excellence Research Chairs (CERC) program (215284), the Natural Sciences and Engineering Research Council of Canada (NSERC), Collaborative Research and Training Experience (CREATE) program and the Canada First Research Excellence Fund (CFREF).

\appendix

\section{} \label{AppendixDD}

\subsection*{Standard theory of dynamical diffraction in a four-blade NI}
 Consider a periodic crystal oriented in the Laue geometry where the transmitted beam and the reflected beam both exit from the same surface of the crystal. Also assume that the crystallographic orientation are perpendicular to the surface of the crystal. The wavefunction of the incoming neutron in free space is represented as a superposition of plane waves
\begin{align}
\ket{\Psi^\text{in}} = \int d\mathbf{k} \,\mu_{\textbf{k}}\, \ket{\mathbf{k}}
\end{align}
 where $\mu_{\mathbf{k}}$ is the probability amplitude of a component $\ket{\mathbf{k}}$ having wavevector $\mathbf{k}={k}_y \hat{e}_y+k_z \hat{e}_z$, see Fig.~\ref{Fig:Laue} for details. 
 According to the standard theory of dynamical diffraction (DD), a component, $\ket{\mathbf{k}}$,  of the wavepacket inside of the crystal generates two excitations along its propagation direction and two along the Bragg diffracted direction.
 At the exit of the blade, the four waves recombine to form the transmitted and reflected which can be expressed as
 \begin{align}
\ket{\Psi}
= \int d\mathbf{k} \ \mu_{\textbf{k}} \Big(t\ket{\mathbf{k}}+r\ket{\mathbf{k}_\text{\tiny H}}\!\Big).\label{Eq:DDOneStandard}
\end{align}
Here, the coefficients $t$ and $r$ are given by ~\cite{Lemmel_2013_ActaCrys,Rauch_2015_Book}
\begin{align}
\begin{split}
t 	&= 	e^{i\chi}e^{-i\frac{\pi z_0}{\Delta_\text{\tiny H}}\eta}\Big( C(\eta)+iS(\eta)\Big),\\
r	&=	-ie^{i\chi}e^{-i\frac{\pi}{\Delta_\text{\tiny H}}(z_0-2z)\eta} \left(\frac{V_\text{\tiny H}}{V_{\text{-\tiny H}}}\right)\ S(\eta),
\label{ddeq2}
\end{split}
\end{align}
where  $z_0$ is the thickness of the crystal whose potential is represented by Fourier components $V_\text{\tiny H}$, the parameter $\Delta_\text{\tiny H}\!=\!{\hbar^2K_\perp\pi}{m^{-1}|V_\text{\tiny H}|^{-1}}$, with $K_\perp\! =\! \sqrt{k_\perp^2-2mV_0/\hbar^2}$, $m$ the mass of the neutron, and the functions
\begin{align*}
C(\eta)& =\cos\left(\frac{\pi z_0}{\Delta_\text{\tiny H}}\sqrt{1+\eta^2}\right),\\ 
S(\eta)&=\frac{\eta}{\sqrt{(1+\eta^2)}}\sin\left(\frac{\pi z_0}{\Delta_\text{\tiny H}}\sqrt{1+\eta^2}\right).
\end{align*}
Finally, $\chi=z_0(K_\perp-k_\perp)$ is the nuclear phase shift due to the crystal which also occurs outside the Bragg condition. The  parameterization of the angular deviation is $\eta=\delta\theta/\Theta_\text{\tiny D}$, where $\Theta_\text{\tiny D}$ is the range of accepted angles by the crystal commonly referred to as the Darwin width. 
The functions $C(\eta)$ and $S(\eta)$ are oscillation functions, with period $\Delta_\text{\tiny H}$. Their oscillation cause the energy to switch between the transmitted and reflected components,  a feature commonly referred to as the Pendel{\"o}sung oscillation~\cite{Rauch_2015_Book}.
 
For convenience in the upcoming sections, we define a function 
\bea
\mathcal{J}_{mn}=\int\! d\eta\ |\mu_\eta|^2|t|^m|r|^n,
\eea
that represents the integrated intensity and the probability of a beam of neutrons undergoing $m$ transmissions and $n$ reflections.

In the case of the four-blade NI in Fig.\ref{Fig:Laue},  if we introduce a control phase difference $\phi$ between the two paths the neutron wavefunction at the output can be expressed as,
\begin{align}
\begin{split}
\ket{\Psi_\text{\tiny NI}} 
= \ket{\Psi^\text{\tiny H}}+\ket{\Psi^\text{\tiny O}},
\end{split}
\end{align}
where, for example, 
\begin{align}
\begin{split}
\ket{\Psi^\text{\tiny H}} &=  \int d\eta \Big( \ket{ \Psi^\text{\tiny I}(\eta)} + e^{i\phi} \ket{ \Psi^\text{\tiny II}(\eta)}\! \Big), \\
&= \int d\eta \mu_{\eta} |tr^3|\Big( e^{i(\varphi_\text{\tiny I}+\varphi_t)} + e^{i\phi}e^{i(\varphi_\text{\tiny II}-\varphi_t)}\Big),
\end{split}
\end{align}
where the blades are assumed to have the same thickness, and  $\varphi_\text{\tiny I}$ and $\varphi_\text{\tiny II}$ are the free space propagation phases of the two components reaching the H-beam via path I and II, and  $\varphi_{t}=\arg [t]$. Similar results can be obtained for the O-beam.

The normalized integrated intensities on detectors placed at the output of the NI and labeled $I_\text{\tiny O}$ for the transmitted direction intensity and $I_\text{\tiny H}$ for the reflected direction intensity can be expressed as
\begin{align}
\begin{split}
I_\text{\tiny H}&= \mathcal{A}_\text{\tiny H}+\mathcal{B}_\text{\tiny H}\cos\left(\phi-\varphi_\text{\tiny H}\right),\\
I_\text{\tiny O}&=  \mathcal{A}_\text{\tiny O}-\mathcal{B}_\text{\tiny H}\cos\left(\phi-\varphi_\text{\tiny H}\right), \label{Eq:FourBlade-DD}
\end{split}
\end{align}
where  $ \mathcal{A}_\text{\tiny O}=\mathcal{J}_{40}+\mathcal{J}_{04}$, $ \mathcal{A}_\text{\tiny H}=2\mathcal{J}_{22}$.
In the four-blade DFS NI with beam displaces in the transverse direction by $\Delta z$, $\mathcal{B}_\text{\tiny H}$ and $\varphi_\text{\tiny H}$ are the the absolute value and the phase of the coherence  function (denoted as $\gamma$ to distinguish from the discrete version $\Gamma$) expressed as,
\begin{align}
\begin{split}
\gamma(\Delta z)&=\frac{1}{\mathcal{N}_0} \int \langle \Psi^\text{\tiny II}(\eta) \ket{ \Psi^\text{\tiny I}(\eta)},\\
&= \int g(\eta) e^{i\left(2\varphi_t- p\Delta z\right)}d\eta,\label{Eq:CohFunction-SDD}
\end{split}
\end{align}
with $\Delta z=2z_{m}-2z_{m^\prime}+z_a-z_s$, $p=2\pi\eta/\Delta_\text{\tiny H}$,  $g(\eta)$ is the angular distribution, and $\mathcal{N}_0$ a normalization constant. The quantities $z_s,z_{m},z_{m^\prime}$, and $z_a$ are the locations of the blades beam splitter, mirrors and analyzer crystals. 

\begin{figure}[t!]
\center
\includegraphics[width=.95\columnwidth]{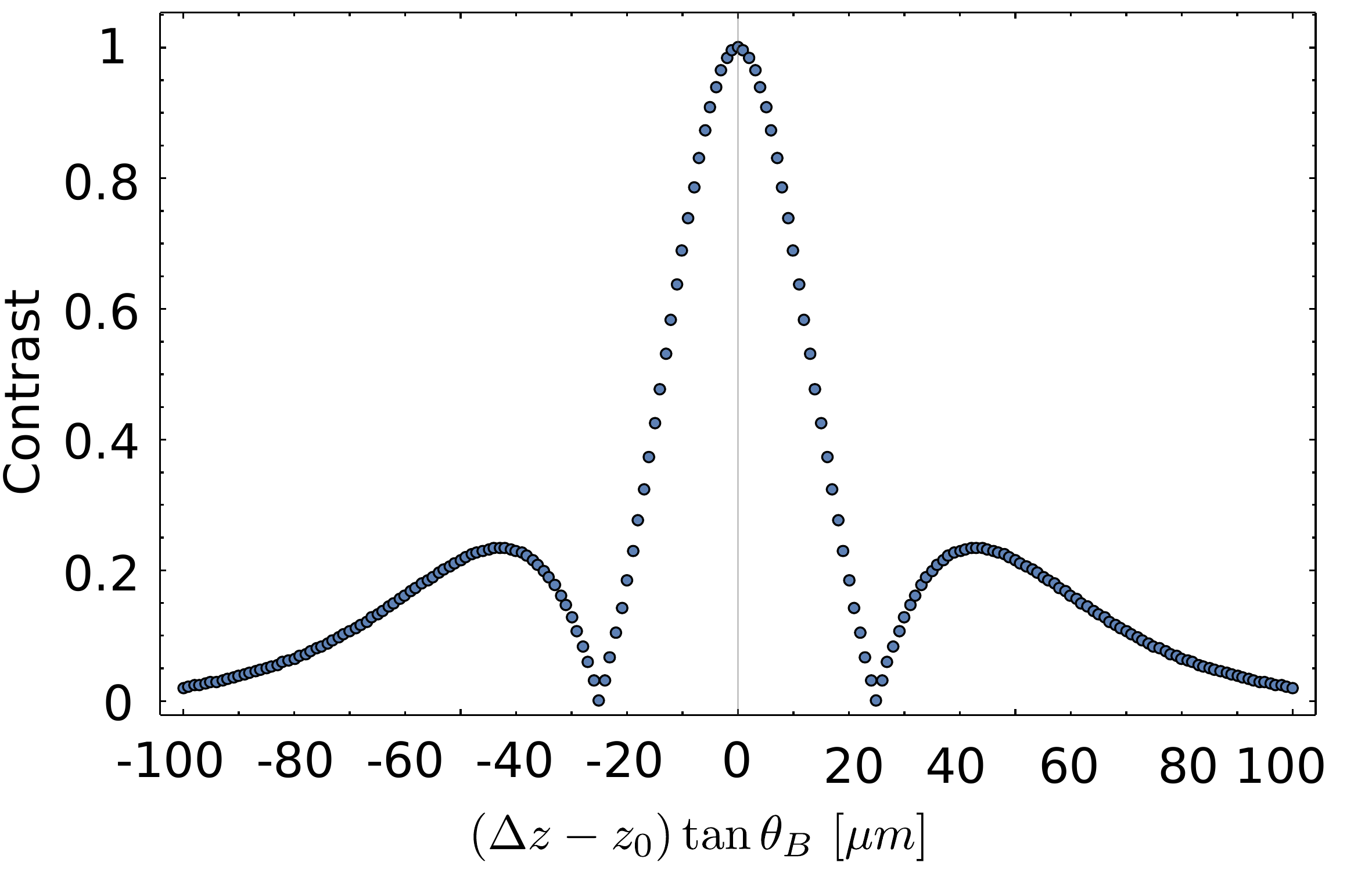}
\caption{ A plot of the contrast as a function of the difference between the  defocussing parameter and the thickness $\Delta z-z_0$. It shows that the contrast can be improved by increasing the defocussing parameter.} \label{Fig:Coherence4BNI-SDD}
\end{figure}

Here we consider the parameters from the four-blade DFS NI design in \cite{Pushin_2009_PRA,Pushin_2011_PRL}. As shown in Eq.~\ref{Eq:CohFunction-SDD}, the focusing condition is $\Delta z=0$ results in a reduction in the contrast for the four-blade DFS NI.
A plot of the contrast or equivalently the absolute value of $\gamma$, versus $\Delta z- z_0$ is shown in Fig.~\ref{Fig:Coherence4BNI-SDD}. The contrast is at a maximum where the NI is defocussed by $\Delta z= z_0$.

\bibliography{NIReference}

\begin{thebibliography}{28}%
\makeatletter
\providecommand \@ifxundefined [1]{%
 \@ifx{#1\undefined}
}%
\providecommand \@ifnum [1]{%
 \ifnum #1\expandafter \@firstoftwo
 \else \expandafter \@secondoftwo
 \fi
}%
\providecommand \@ifx [1]{%
 \ifx #1\expandafter \@firstoftwo
 \else \expandafter \@secondoftwo
 \fi
}%
\providecommand \natexlab [1]{#1}%
\providecommand \enquote  [1]{``#1''}%
\providecommand \bibnamefont  [1]{#1}%
\providecommand \bibfnamefont [1]{#1}%
\providecommand \citenamefont [1]{#1}%
\providecommand \href@noop [0]{\@secondoftwo}%
\providecommand \href [0]{\begingroup \@sanitize@url \@href}%
\providecommand \@href[1]{\@@startlink{#1}\@@href}%
\providecommand \@@href[1]{\endgroup#1\@@endlink}%
\providecommand \@sanitize@url [0]{\catcode `\\12\catcode `\$12\catcode
  `\&12\catcode `\#12\catcode `\^12\catcode `\_12\catcode `\%12\relax}%
\providecommand \@@startlink[1]{}%
\providecommand \@@endlink[0]{}%
\providecommand \url  [0]{\begingroup\@sanitize@url \@url }%
\providecommand \@url [1]{\endgroup\@href {#1}{\urlprefix }}%
\providecommand \urlprefix  [0]{URL }%
\providecommand \Eprint [0]{\href }%
\providecommand \doibase [0]{http://dx.doi.org/}%
\providecommand \selectlanguage [0]{\@gobble}%
\providecommand \bibinfo  [0]{\@secondoftwo}%
\providecommand \bibfield  [0]{\@secondoftwo}%
\providecommand \translation [1]{[#1]}%
\providecommand \BibitemOpen [0]{}%
\providecommand \bibitemStop [0]{}%
\providecommand \bibitemNoStop [0]{.\EOS\space}%
\providecommand \EOS [0]{\spacefactor3000\relax}%
\providecommand \BibitemShut  [1]{\csname bibitem#1\endcsname}%
\let\auto@bib@innerbib\@empty
\bibitem [{\citenamefont {Rauch}\ \emph {et~al.}(1975)\citenamefont {Rauch},
  \citenamefont {Zeilinger}, \citenamefont {Badurek}, \citenamefont {Wilfing},
  \citenamefont {Bauspiess},\ and\ \citenamefont
  {Bonse}}]{Rauch_1975_PhysLetta}%
  \BibitemOpen
  \bibfield  {author} {\bibinfo {author} {\bibfnamefont {H.}~\bibnamefont
  {Rauch}}, \bibinfo {author} {\bibfnamefont {A.}~\bibnamefont {Zeilinger}},
  \bibinfo {author} {\bibfnamefont {G.}~\bibnamefont {Badurek}}, \bibinfo
  {author} {\bibfnamefont {A.}~\bibnamefont {Wilfing}}, \bibinfo {author}
  {\bibfnamefont {W.}~\bibnamefont {Bauspiess}}, \ and\ \bibinfo {author}
  {\bibfnamefont {U.}~\bibnamefont {Bonse}},\ }\bibfield  {title} {\enquote
  {\bibinfo {title} {Verification of coherent spinor rotation of fermions},}\
  }\href {\doibase DOI: 10.1016/0375-9601(75)90798-7} {\bibfield  {journal}
  {\bibinfo  {journal} {Phys. Lett.}\ }\textbf {\bibinfo {volume} {A54}},\
  \bibinfo {pages} {425 -- 427} (\bibinfo {year} {1975})}\BibitemShut {NoStop}%
\bibitem [{\citenamefont {Colella}\ \emph {et~al.}(1975)\citenamefont
  {Colella}, \citenamefont {Overhauser},\ and\ \citenamefont
  {Werner}}]{Colella_1975_PRL}%
  \BibitemOpen
  \bibfield  {author} {\bibinfo {author} {\bibfnamefont {R.}~\bibnamefont
  {Colella}}, \bibinfo {author} {\bibfnamefont {A.~W.}\ \bibnamefont
  {Overhauser}}, \ and\ \bibinfo {author} {\bibfnamefont {S.~A.}\ \bibnamefont
  {Werner}},\ }\bibfield  {title} {\enquote {\bibinfo {title} {Observation of
  gravitationally induced quantum interference},}\ }\href {\doibase
  10.1103/PhysRevLett.34.1472} {\bibfield  {journal} {\bibinfo  {journal}
  {Phys. Rev. Lett.}\ }\textbf {\bibinfo {volume} {34}},\ \bibinfo {pages}
  {1472--1474} (\bibinfo {year} {1975})}\BibitemShut {NoStop}%
\bibitem [{\citenamefont {Schoen}\ \emph {et~al.}(2003)\citenamefont {Schoen},
  \citenamefont {Jacobson}, \citenamefont {Arif}, \citenamefont {Huffman},
  \citenamefont {Black}, \citenamefont {Snow}, \citenamefont {Lamoreaux},
  \citenamefont {Kaiser},\ and\ \citenamefont {Werner}}]{Schoen_2003_PhysRev}%
  \BibitemOpen
  \bibfield  {author} {\bibinfo {author} {\bibfnamefont {K.}~\bibnamefont
  {Schoen}}, \bibinfo {author} {\bibfnamefont {D.~L.}\ \bibnamefont
  {Jacobson}}, \bibinfo {author} {\bibfnamefont {M.}~\bibnamefont {Arif}},
  \bibinfo {author} {\bibfnamefont {P.~R.}\ \bibnamefont {Huffman}}, \bibinfo
  {author} {\bibfnamefont {T.~C.}\ \bibnamefont {Black}}, \bibinfo {author}
  {\bibfnamefont {W.~M.}\ \bibnamefont {Snow}}, \bibinfo {author}
  {\bibfnamefont {S.~K.}\ \bibnamefont {Lamoreaux}}, \bibinfo {author}
  {\bibfnamefont {H.}~\bibnamefont {Kaiser}}, \ and\ \bibinfo {author}
  {\bibfnamefont {S.~A.}\ \bibnamefont {Werner}},\ }\bibfield  {title}
  {\enquote {\bibinfo {title} {Precision neutron interferometric measurements
  and updated evaluations of the $n-p$ and $n-d$ coherent neutron scattering
  lengths},}\ }\href {\doibase 10.1103/PhysRevC.67.044005} {\bibfield
  {journal} {\bibinfo  {journal} {Phys. Rev.}\ }\textbf {\bibinfo {volume}
  {C67}},\ \bibinfo {pages} {044005} (\bibinfo {year} {2003})}\BibitemShut
  {NoStop}%
\bibitem [{\citenamefont {Greene}\ and\ \citenamefont
  {Gudkov}(2007)}]{Green_2007_PRC}%
  \BibitemOpen
  \bibfield  {author} {\bibinfo {author} {\bibfnamefont {Geoffrey~L.}\
  \bibnamefont {Greene}}\ and\ \bibinfo {author} {\bibfnamefont {Vladimir}\
  \bibnamefont {Gudkov}},\ }\bibfield  {title} {\enquote {\bibinfo {title}
  {Neutron interferometric method to provide improved constraints on
  non-newtonian gravity at the nanometer scale},}\ }\href {\doibase
  10.1103/PhysRevC.75.015501} {\bibfield  {journal} {\bibinfo  {journal} {Phys.
  Rev. C}\ }\textbf {\bibinfo {volume} {75}},\ \bibinfo {pages} {015501}
  (\bibinfo {year} {2007})}\BibitemShut {NoStop}%
\bibitem [{\citenamefont {Ioffe}\ and\ \citenamefont
  {Vrana}(2002)}]{Ioffe_2002_ApplPhysA}%
  \BibitemOpen
  \bibfield  {author} {\bibinfo {author} {\bibfnamefont {A.}~\bibnamefont
  {Ioffe}}\ and\ \bibinfo {author} {\bibfnamefont {M.}~\bibnamefont {Vrana}},\
  }\bibfield  {title} {\enquote {\bibinfo {title} {A new neutron interferometry
  approach in the determination of the neutron--electron interaction
  amplitude},}\ }\href {\doibase 10.1007/s003390101097} {\bibfield  {journal}
  {\bibinfo  {journal} {Applied Physics A}\ }\textbf {\bibinfo {volume} {74}},\
  \bibinfo {pages} {s314--s316} (\bibinfo {year} {2002})}\BibitemShut {NoStop}%
\bibitem [{\citenamefont {Rauch}\ and\ \citenamefont
  {Werner}(2015)}]{Rauch_2015_Book}%
  \BibitemOpen
  \bibfield  {author} {\bibinfo {author} {\bibfnamefont {H.}~\bibnamefont
  {Rauch}}\ and\ \bibinfo {author} {\bibfnamefont {S.~A.}\ \bibnamefont
  {Werner}},\ }\href@noop {} {\emph {\bibinfo {title} {Neutron Interferometry:
  Lessons in Experimental Quantum Mechanics, Wave-Particle Duality, and
  Entanglement}}},\ \bibinfo {edition} {2nd}\ ed.\ (\bibinfo  {publisher}
  {Oxford University Press},\ \bibinfo {year} {2015})\BibitemShut {NoStop}%
\bibitem [{\citenamefont {Clark}\ \emph {et~al.}(2015)\citenamefont {Clark},
  \citenamefont {Barankov}, \citenamefont {Huber}, \citenamefont {Arif},
  \citenamefont {Cory},\ and\ \citenamefont {Pushin}}]{oam}%
  \BibitemOpen
  \bibfield  {author} {\bibinfo {author} {\bibfnamefont {Charles~W}\
  \bibnamefont {Clark}}, \bibinfo {author} {\bibfnamefont {Roman}\ \bibnamefont
  {Barankov}}, \bibinfo {author} {\bibfnamefont {Michael~G}\ \bibnamefont
  {Huber}}, \bibinfo {author} {\bibfnamefont {Muhammad}\ \bibnamefont {Arif}},
  \bibinfo {author} {\bibfnamefont {David~G}\ \bibnamefont {Cory}}, \ and\
  \bibinfo {author} {\bibfnamefont {Dmitry~A}\ \bibnamefont {Pushin}},\
  }\bibfield  {title} {\enquote {\bibinfo {title} {Controlling neutron orbital
  angular momentum},}\ }\href@noop {} {\bibfield  {journal} {\bibinfo
  {journal} {Nature}\ }\textbf {\bibinfo {volume} {525}},\ \bibinfo {pages}
  {504--506} (\bibinfo {year} {2015})}\BibitemShut {NoStop}%
\bibitem [{\citenamefont {Li}\ \emph {et~al.}(2016)\citenamefont {Li},
  \citenamefont {Arif}, \citenamefont {Cory}, \citenamefont {Haun},
  \citenamefont {Heacock}, \citenamefont {Huber}, \citenamefont {Nsofini},
  \citenamefont {Pushin}, \citenamefont {Saggu}, \citenamefont {Sarenac} \emph
  {et~al.}}]{chameleon}%
  \BibitemOpen
  \bibfield  {author} {\bibinfo {author} {\bibfnamefont {K}~\bibnamefont {Li}},
  \bibinfo {author} {\bibfnamefont {M}~\bibnamefont {Arif}}, \bibinfo {author}
  {\bibfnamefont {DG}~\bibnamefont {Cory}}, \bibinfo {author} {\bibfnamefont
  {R}~\bibnamefont {Haun}}, \bibinfo {author} {\bibfnamefont {B}~\bibnamefont
  {Heacock}}, \bibinfo {author} {\bibfnamefont {MG}~\bibnamefont {Huber}},
  \bibinfo {author} {\bibfnamefont {J}~\bibnamefont {Nsofini}}, \bibinfo
  {author} {\bibfnamefont {DA}~\bibnamefont {Pushin}}, \bibinfo {author}
  {\bibfnamefont {P}~\bibnamefont {Saggu}}, \bibinfo {author} {\bibfnamefont
  {D}~\bibnamefont {Sarenac}},  \emph {et~al.},\ }\bibfield  {title} {\enquote
  {\bibinfo {title} {Neutron limit on the strongly-coupled chameleon field},}\
  }\href@noop {} {\bibfield  {journal} {\bibinfo  {journal} {Physical Review
  D}\ }\textbf {\bibinfo {volume} {93}},\ \bibinfo {pages} {062001} (\bibinfo
  {year} {2016})}\BibitemShut {NoStop}%
\bibitem [{\citenamefont {Sarenac}\ \emph {et~al.}(2016)\citenamefont
  {Sarenac}, \citenamefont {Huber}, \citenamefont {Heacock}, \citenamefont
  {Arif}, \citenamefont {Clark}, \citenamefont {Cory}, \citenamefont {Shahi},\
  and\ \citenamefont {Pushin}}]{holography}%
  \BibitemOpen
  \bibfield  {author} {\bibinfo {author} {\bibfnamefont {Dusan}\ \bibnamefont
  {Sarenac}}, \bibinfo {author} {\bibfnamefont {Michael~G}\ \bibnamefont
  {Huber}}, \bibinfo {author} {\bibfnamefont {Benjamin}\ \bibnamefont
  {Heacock}}, \bibinfo {author} {\bibfnamefont {Muhammad}\ \bibnamefont
  {Arif}}, \bibinfo {author} {\bibfnamefont {Charles~W}\ \bibnamefont {Clark}},
  \bibinfo {author} {\bibfnamefont {David~G}\ \bibnamefont {Cory}}, \bibinfo
  {author} {\bibfnamefont {Chandra~B}\ \bibnamefont {Shahi}}, \ and\ \bibinfo
  {author} {\bibfnamefont {Dmitry~A}\ \bibnamefont {Pushin}},\ }\bibfield
  {title} {\enquote {\bibinfo {title} {Holography with a neutron
  interferometer},}\ }\href@noop {} {\bibfield  {journal} {\bibinfo  {journal}
  {Optics Express}\ }\textbf {\bibinfo {volume} {24}},\ \bibinfo {pages}
  {22528--22535} (\bibinfo {year} {2016})}\BibitemShut {NoStop}%
\bibitem [{\citenamefont {Lemmel}\ \emph {et~al.}(2015)\citenamefont {Lemmel},
  \citenamefont {Brax}, \citenamefont {Ivanov}, \citenamefont {Jenke},
  \citenamefont {Pignol}, \citenamefont {Pitschmann}, \citenamefont {Potocar},
  \citenamefont {Wellenzohn}, \citenamefont {Zawisky},\ and\ \citenamefont
  {Abele}}]{lemmel2015neutron}%
  \BibitemOpen
  \bibfield  {author} {\bibinfo {author} {\bibfnamefont {H}~\bibnamefont
  {Lemmel}}, \bibinfo {author} {\bibfnamefont {Ph}~\bibnamefont {Brax}},
  \bibinfo {author} {\bibfnamefont {AN}~\bibnamefont {Ivanov}}, \bibinfo
  {author} {\bibfnamefont {T}~\bibnamefont {Jenke}}, \bibinfo {author}
  {\bibfnamefont {G}~\bibnamefont {Pignol}}, \bibinfo {author} {\bibfnamefont
  {M}~\bibnamefont {Pitschmann}}, \bibinfo {author} {\bibfnamefont
  {T}~\bibnamefont {Potocar}}, \bibinfo {author} {\bibfnamefont
  {M}~\bibnamefont {Wellenzohn}}, \bibinfo {author} {\bibfnamefont
  {M}~\bibnamefont {Zawisky}}, \ and\ \bibinfo {author} {\bibfnamefont
  {H}~\bibnamefont {Abele}},\ }\bibfield  {title} {\enquote {\bibinfo {title}
  {Neutron interferometry constrains dark energy chameleon fields},}\
  }\href@noop {} {\bibfield  {journal} {\bibinfo  {journal} {Physics Letters
  B}\ }\textbf {\bibinfo {volume} {743}},\ \bibinfo {pages} {310--314}
  (\bibinfo {year} {2015})}\BibitemShut {NoStop}%
\bibitem [{\citenamefont {Shahi}\ \emph {et~al.}(2016)\citenamefont {Shahi},
  \citenamefont {Arif}, \citenamefont {Cory}, \citenamefont {Mineeva},
  \citenamefont {Nsofini}, \citenamefont {Sarenac}, \citenamefont {Williams},
  \citenamefont {Huber},\ and\ \citenamefont {Pushin}}]{Shahi_2016_NIM}%
  \BibitemOpen
  \bibfield  {author} {\bibinfo {author} {\bibfnamefont {C.B.}\ \bibnamefont
  {Shahi}}, \bibinfo {author} {\bibfnamefont {M.}~\bibnamefont {Arif}},
  \bibinfo {author} {\bibfnamefont {D.G.}\ \bibnamefont {Cory}}, \bibinfo
  {author} {\bibfnamefont {T.}~\bibnamefont {Mineeva}}, \bibinfo {author}
  {\bibfnamefont {J.}~\bibnamefont {Nsofini}}, \bibinfo {author} {\bibfnamefont
  {D.}~\bibnamefont {Sarenac}}, \bibinfo {author} {\bibfnamefont {C.J.}\
  \bibnamefont {Williams}}, \bibinfo {author} {\bibfnamefont {M.G.}\
  \bibnamefont {Huber}}, \ and\ \bibinfo {author} {\bibfnamefont {D.A.}\
  \bibnamefont {Pushin}},\ }\bibfield  {title} {\enquote {\bibinfo {title} {A
  new polarized neutron interferometry facility at the ncnr},}\ }\href
  {\doibase http://dx.doi.org/10.1016/j.nima.2016.01.023} {\bibfield  {journal}
  {\bibinfo  {journal} {Nuclear Instruments and Methods in Physics Research
  Section A}\ }\textbf {\bibinfo {volume} {813}},\ \bibinfo {pages} {111 --
  122} (\bibinfo {year} {2016})}\BibitemShut {NoStop}%
\bibitem [{\citenamefont {Pushin}\ \emph {et~al.}(2015)\citenamefont {Pushin},
  \citenamefont {Huber}, \citenamefont {Arif}, \citenamefont {Shahi},
  \citenamefont {Nsofini}, \citenamefont {Wood}, \citenamefont {Sarenac},\ and\
  \citenamefont {Cory}}]{Pushin_2015_AHEP}%
  \BibitemOpen
  \bibfield  {author} {\bibinfo {author} {\bibfnamefont {D.~A.}\ \bibnamefont
  {Pushin}}, \bibinfo {author} {\bibfnamefont {M.~G.}\ \bibnamefont {Huber}},
  \bibinfo {author} {\bibfnamefont {M.}~\bibnamefont {Arif}}, \bibinfo {author}
  {\bibfnamefont {C.~B.}\ \bibnamefont {Shahi}}, \bibinfo {author}
  {\bibfnamefont {J.}~\bibnamefont {Nsofini}}, \bibinfo {author} {\bibfnamefont
  {C.~J.}\ \bibnamefont {Wood}}, \bibinfo {author} {\bibfnamefont
  {D.}~\bibnamefont {Sarenac}}, \ and\ \bibinfo {author} {\bibfnamefont
  {D.~G.}\ \bibnamefont {Cory}},\ }\bibfield  {title} {\enquote {\bibinfo
  {title} {Neutron interferometry at the national institute of standards and
  technology},}\ }\href {\doibase 10.1155/2015/687480} {\bibfield  {journal}
  {\bibinfo  {journal} {Advances in High Energy Physics}\ } (\bibinfo {year}
  {2015}),\ 10.1155/2015/687480}\BibitemShut {NoStop}%
\bibitem [{\citenamefont {Saggu}\ \emph {et~al.}(2016)\citenamefont {Saggu},
  \citenamefont {Mineeva}, \citenamefont {Arif}, \citenamefont {Cory},
  \citenamefont {Haun}, \citenamefont {Heacock}, \citenamefont {Huber},
  \citenamefont {Li}, \citenamefont {Nsofini}, \citenamefont {Sarenac} \emph
  {et~al.}}]{saggu2016decoupling}%
  \BibitemOpen
  \bibfield  {author} {\bibinfo {author} {\bibfnamefont {Parminder}\
  \bibnamefont {Saggu}}, \bibinfo {author} {\bibfnamefont {Taisiya}\
  \bibnamefont {Mineeva}}, \bibinfo {author} {\bibfnamefont {Muhammad}\
  \bibnamefont {Arif}}, \bibinfo {author} {\bibfnamefont {DG}~\bibnamefont
  {Cory}}, \bibinfo {author} {\bibfnamefont {Robert}\ \bibnamefont {Haun}},
  \bibinfo {author} {\bibfnamefont {Ben}\ \bibnamefont {Heacock}}, \bibinfo
  {author} {\bibfnamefont {MG}~\bibnamefont {Huber}}, \bibinfo {author}
  {\bibfnamefont {Ke}~\bibnamefont {Li}}, \bibinfo {author} {\bibfnamefont
  {Joachim}\ \bibnamefont {Nsofini}}, \bibinfo {author} {\bibfnamefont {Dusan}\
  \bibnamefont {Sarenac}},  \emph {et~al.},\ }\bibfield  {title} {\enquote
  {\bibinfo {title} {Decoupling of a neutron interferometer from temperature
  gradients},}\ }\href@noop {} {\bibfield  {journal} {\bibinfo  {journal}
  {Review of Scientific Instruments}\ }\textbf {\bibinfo {volume} {87}},\
  \bibinfo {pages} {123507} (\bibinfo {year} {2016})}\BibitemShut {NoStop}%
\bibitem [{\citenamefont {Pushin}\ \emph {et~al.}(2011)\citenamefont {Pushin},
  \citenamefont {Huber}, \citenamefont {Arif},\ and\ \citenamefont
  {Cory}}]{Pushin_2011_PRL}%
  \BibitemOpen
  \bibfield  {author} {\bibinfo {author} {\bibfnamefont {D.A.}\ \bibnamefont
  {Pushin}}, \bibinfo {author} {\bibfnamefont {M.G.}\ \bibnamefont {Huber}},
  \bibinfo {author} {\bibfnamefont {M.}~\bibnamefont {Arif}}, \ and\ \bibinfo
  {author} {\bibfnamefont {D.G.}\ \bibnamefont {Cory}},\ }\bibfield  {title}
  {\enquote {\bibinfo {title} {Experimental realization of decoherence-free
  subspace in neutron interferometry},}\ }\href {\doibase
  10.1103/PhysRevLett.107.150401} {\bibfield  {journal} {\bibinfo  {journal}
  {Phys. Rev. Lett.}\ }\textbf {\bibinfo {volume} {107}},\ \bibinfo {pages}
  {150401} (\bibinfo {year} {2011})}\BibitemShut {NoStop}%
\bibitem [{\citenamefont {Pushin}\ \emph {et~al.}(2009)\citenamefont {Pushin},
  \citenamefont {Arif},\ and\ \citenamefont {Cory}}]{Pushin_2009_PRA}%
  \BibitemOpen
  \bibfield  {author} {\bibinfo {author} {\bibfnamefont {D.~A.}\ \bibnamefont
  {Pushin}}, \bibinfo {author} {\bibfnamefont {M.}~\bibnamefont {Arif}}, \ and\
  \bibinfo {author} {\bibfnamefont {D.~G.}\ \bibnamefont {Cory}},\ }\bibfield
  {title} {\enquote {\bibinfo {title} {Decoherence-free neutron
  interferometry},}\ }\href {\doibase 10.1103/PhysRevA.79.053635} {\bibfield
  {journal} {\bibinfo  {journal} {Phys Rev}\ }\textbf {\bibinfo {volume}
  {79}},\ \bibinfo {eid} {053635} (\bibinfo {year} {2009})}\BibitemShut
  {NoStop}%
\bibitem [{\citenamefont {Nsofini}\ \emph {et~al.}(2017)\citenamefont
  {Nsofini}, \citenamefont {Sarenac}, \citenamefont {Ghofrani}, \citenamefont
  {Huber}, \citenamefont {Arif}, \citenamefont {Cory},\ and\ \citenamefont
  {Pushin}}]{Nsofini_2017_JAP}%
  \BibitemOpen
  \bibfield  {author} {\bibinfo {author} {\bibfnamefont {J.}~\bibnamefont
  {Nsofini}}, \bibinfo {author} {\bibfnamefont {D.}~\bibnamefont {Sarenac}},
  \bibinfo {author} {\bibfnamefont {K.}~\bibnamefont {Ghofrani}}, \bibinfo
  {author} {\bibfnamefont {M.~G.}\ \bibnamefont {Huber}}, \bibinfo {author}
  {\bibfnamefont {M.}~\bibnamefont {Arif}}, \bibinfo {author} {\bibfnamefont
  {D.~G.}\ \bibnamefont {Cory}}, \ and\ \bibinfo {author} {\bibfnamefont
  {D.~A.}\ \bibnamefont {Pushin}},\ }\bibfield  {title} {\enquote {\bibinfo
  {title} {Noise refocusing in a five-blade neutron interferometer},}\ }\href
  {\doibase 10.1063/1.4996866} {\bibfield  {journal} {\bibinfo  {journal}
  {Journal of Applied Physics}\ }\textbf {\bibinfo {volume} {122}},\ \bibinfo
  {pages} {054501} (\bibinfo {year} {2017})},\ \Eprint
  {http://arxiv.org/abs/https://doi.org/10.1063/1.4996866}
  {https://doi.org/10.1063/1.4996866} \BibitemShut {NoStop}%
\bibitem [{\citenamefont {Nsofini}\ \emph {et~al.}(2016)\citenamefont
  {Nsofini}, \citenamefont {Ghofrani}, \citenamefont {Sarenac}, \citenamefont
  {Cory},\ and\ \citenamefont {Pushin}}]{Nsofini_2016_PhysRevA}%
  \BibitemOpen
  \bibfield  {author} {\bibinfo {author} {\bibfnamefont {J.}~\bibnamefont
  {Nsofini}}, \bibinfo {author} {\bibfnamefont {K.}~\bibnamefont {Ghofrani}},
  \bibinfo {author} {\bibfnamefont {D.}~\bibnamefont {Sarenac}}, \bibinfo
  {author} {\bibfnamefont {D.~G.}\ \bibnamefont {Cory}}, \ and\ \bibinfo
  {author} {\bibfnamefont {D.~A.}\ \bibnamefont {Pushin}},\ }\bibfield  {title}
  {\enquote {\bibinfo {title} {Quantum-information approach to dynamical
  diffraction theory},}\ }\href {\doibase 10.1103/PhysRevA.94.062311}
  {\bibfield  {journal} {\bibinfo  {journal} {Phys. Rev. A}\ }\textbf {\bibinfo
  {volume} {94}},\ \bibinfo {pages} {062311} (\bibinfo {year}
  {2016})}\BibitemShut {NoStop}%
\bibitem [{\citenamefont {Authier}(2006)}]{Authier_2006_Book}%
  \BibitemOpen
  \bibfield  {author} {\bibinfo {author} {\bibfnamefont {A.}~\bibnamefont
  {Authier}},\ }\href@noop {} {\emph {\bibinfo {title} {Dynamical Theory of
  X-ray Diffraction}}}\ (\bibinfo  {publisher} {Oxford},\ \bibinfo {address}
  {Oxford Univ Press},\ \bibinfo {year} {2006})\BibitemShut {NoStop}%
\bibitem [{\citenamefont {Utsuro}\ and\ \citenamefont
  {Ignatovich}(2012)}]{Utsuro_2012_Book}%
  \BibitemOpen
  \bibfield  {author} {\bibinfo {author} {\bibfnamefont {Masahiko}\
  \bibnamefont {Utsuro}}\ and\ \bibinfo {author} {\bibfnamefont {Vladimir~K.}\
  \bibnamefont {Ignatovich}},\ }\href@noop {} {\emph {\bibinfo {title}
  {Handbook of Neutron Optics}}}\ (\bibinfo  {publisher} {Wiley-VCH},\ \bibinfo
  {year} {2012})\BibitemShut {NoStop}%
\bibitem [{\citenamefont {Zachariasen}(1945)}]{Zachariasen_1945_Book}%
  \BibitemOpen
  \bibfield  {author} {\bibinfo {author} {\bibfnamefont {W.~H.}\ \bibnamefont
  {Zachariasen}},\ }\href@noop {} {\emph {\bibinfo {title} {Theory of X-ray
  Diffraction in Crystals}}}\ (\bibinfo  {publisher} {Wiley},\ \bibinfo
  {address} {New York},\ \bibinfo {year} {1945})\BibitemShut {NoStop}%
\bibitem [{\citenamefont {Batterman}\ and\ \citenamefont
  {Cole}(1964)}]{Batterman_1964_RevModPhys}%
  \BibitemOpen
  \bibfield  {author} {\bibinfo {author} {\bibfnamefont {B.~W.}\ \bibnamefont
  {Batterman}}\ and\ \bibinfo {author} {\bibfnamefont {H.}~\bibnamefont
  {Cole}},\ }\bibfield  {title} {\enquote {\bibinfo {title} {Dynamical
  diffraction of x rays by perfect crystals},}\ }\href {\doibase
  10.1103/RevModPhys.36.681} {\bibfield  {journal} {\bibinfo  {journal} {Rev.
  Mod. Phys.}\ }\textbf {\bibinfo {volume} {36}},\ \bibinfo {pages} {681--717}
  (\bibinfo {year} {1964})}\BibitemShut {NoStop}%
\bibitem [{\citenamefont {Sears}(1989)}]{Sears_1989_Book}%
  \BibitemOpen
  \bibfield  {author} {\bibinfo {author} {\bibfnamefont {Varley~F.}\
  \bibnamefont {Sears}},\ }\href@noop {} {\emph {\bibinfo {title} {An
  Introduction to the Theory of Neutron Optical Phenomena and their
  Applications}}}\ (\bibinfo  {publisher} {Oxford University Press},\ \bibinfo
  {address} {New York},\ \bibinfo {year} {1989})\BibitemShut {NoStop}%
\bibitem [{\citenamefont {Pushin}(2006)}]{pushin2006coherent}%
  \BibitemOpen
  \bibfield  {author} {\bibinfo {author} {\bibfnamefont {Dmitry~A}\
  \bibnamefont {Pushin}},\ }\emph {\bibinfo {title} {Coherent control of
  neutron interferometry}},\ \href@noop {} {Ph.D. thesis},\ \bibinfo  {school}
  {Massachusetts Institute of Technology} (\bibinfo {year} {2006})\BibitemShut
  {NoStop}%
\bibitem [{\citenamefont {Petrascheck}(1988)}]{PETRASCHECK_1988_PhysicaB}%
  \BibitemOpen
  \bibfield  {author} {\bibinfo {author} {\bibfnamefont {D.}~\bibnamefont
  {Petrascheck}},\ }\bibfield  {title} {\enquote {\bibinfo {title} {On
  coherence in crystal optics},}\ }\href {\doibase
  https://doi.org/10.1016/0378-4363(88)90162-3} {\bibfield  {journal} {\bibinfo
   {journal} {Physica B+C}\ }\textbf {\bibinfo {volume} {151}},\ \bibinfo
  {pages} {171 -- 175} (\bibinfo {year} {1988})}\BibitemShut {NoStop}%
\bibitem [{\citenamefont {Petrascheck}\ and\ \citenamefont
  {Folk}(1976)}]{Petrascheck_1976_PhysStatSol}%
  \BibitemOpen
  \bibfield  {author} {\bibinfo {author} {\bibfnamefont {D.}~\bibnamefont
  {Petrascheck}}\ and\ \bibinfo {author} {\bibfnamefont {R.}~\bibnamefont
  {Folk}},\ }\bibfield  {title} {\enquote {\bibinfo {title} {Theory of
  symmetric lll interferometer with arbitrary absoption.}}\ }\href@noop {}
  {\bibfield  {journal} {\bibinfo  {journal} {Phys. Stat. Sol.}\ }\textbf
  {\bibinfo {volume} {34}},\ \bibinfo {pages} {147} (\bibinfo {year}
  {1976})}\BibitemShut {NoStop}%
\bibitem [{\citenamefont {Petrascheck}\ and\ \citenamefont
  {Rauch}(1984)}]{Petrascheck_1984_ActaCrys}%
  \BibitemOpen
  \bibfield  {author} {\bibinfo {author} {\bibfnamefont {D.}~\bibnamefont
  {Petrascheck}}\ and\ \bibinfo {author} {\bibfnamefont {H.}~\bibnamefont
  {Rauch}},\ }\bibfield  {title} {\enquote {\bibinfo {title} {Multiple laue
  rocking curves},}\ }\href@noop {} {\bibfield  {journal} {\bibinfo  {journal}
  {Acta Cryst.}\ }\textbf {\bibinfo {volume} {A40}},\ \bibinfo {pages}
  {445--450} (\bibinfo {year} {1984})}\BibitemShut {NoStop}%
\bibitem [{\citenamefont {Pushin}\ \emph {et~al.}(2008)\citenamefont {Pushin},
  \citenamefont {Arif}, \citenamefont {Huber},\ and\ \citenamefont
  {Cory}}]{Pushin_2008_PRL}%
  \BibitemOpen
  \bibfield  {author} {\bibinfo {author} {\bibfnamefont {D.~A.}\ \bibnamefont
  {Pushin}}, \bibinfo {author} {\bibfnamefont {M.}~\bibnamefont {Arif}},
  \bibinfo {author} {\bibfnamefont {M.~G.}\ \bibnamefont {Huber}}, \ and\
  \bibinfo {author} {\bibfnamefont {D.~G.}\ \bibnamefont {Cory}},\ }\bibfield
  {title} {\enquote {\bibinfo {title} {Measurements of the vertical coherence
  length in neutron interferometry},}\ }\href {\doibase
  10.1103/PhysRevLett.100.250404} {\bibfield  {journal} {\bibinfo  {journal}
  {Phys. Rev. Lett.}\ }\textbf {\bibinfo {volume} {100}},\ \bibinfo {eid}
  {250404} (\bibinfo {year} {2008})}\BibitemShut {NoStop}%
\bibitem [{\citenamefont {Lemmel}(2013)}]{Lemmel_2013_ActaCrys}%
  \BibitemOpen
  \bibfield  {author} {\bibinfo {author} {\bibfnamefont {H.}~\bibnamefont
  {Lemmel}},\ }\bibfield  {title} {\enquote {\bibinfo {title} {{Influence of
  Bragg diffraction on perfect crystal neutron phase shifters and the exact
  solution of the two-beam case in the dynamical diffraction theory}},}\ }\href
  {\doibase 10.1107/S0108767313014293} {\bibfield  {journal} {\bibinfo
  {journal} {Acta Crystallographica Section A}\ }\textbf {\bibinfo {volume}
  {69}},\ \bibinfo {pages} {459--474} (\bibinfo {year} {2013})}\BibitemShut
  {NoStop}%
\end{thebibliography}%

\end{document}